\documentclass[aps,pra,reprint,longbibliography]{revtex4-1}

\usepackage{graphicx}
\usepackage{physics}

\begin{document}

\title{Characterization of an atom interferometer in the quasi-Bragg regime.}
\author{A. B\'eguin} 
\author{T. Rodzinka}
\author{J. Vigu\'e}
\author{B. Allard}
\author{A. Gauguet}
\email{gauguet@irsamc.ups-tlse.fr}

\affiliation{Laboratoire Collisions Agr\'egats R\'eactivit\'e, Universit\'e Paul Sabatier, 118 route de Narbonne, 31062 Toulouse Cedex 4, France}

\begin{abstract}
We provide a comprehensive study of ultra-cold atom diffraction by an optical lattice. We focus on an intermediate regime between the Raman-Nath and the Bragg regimes, the so-called quasi-Bragg regime. The experimental results are in a good agreement with a full numerical integration of the Schr\"odinger equation. We investigate the long pulse regime limited by a strong velocity selection and the short pulse regime limited by "non-adiabatic losses". For each of these regimes, we estimate the multi-port features of the Bragg interferometers. Finally, we discuss the best compromise between these two regimes, considering the diffraction phase shift and the existence of parasitic interferometers.
\end{abstract}

\maketitle

%%%%%%%%%%%%%%%%%%%%%%%%%%%%%%%
%%%%%%%%%%%%%%%%%%%%%%%%%%%%%%%
\section{Introduction}
\label{intro}
In 1933, \citet{kapitza_dirac_1933} proposed the diffraction of electrons by a standing light wave for studying the stimulated emission process. Because of the weakness of the interaction between free electrons and photons, the first convincing experimental observation was not performed until 2001 \cite{Freimund2001}. However, an atomic analog of the Kapitza-Dirac effect was proposed by \citet{Altshuler66} who pointed out the huge enhancement of the diffraction probability if one replaces electrons by atoms and if one uses quasi-resonant laser radiation. Following this idea, the atomic diffraction by a standing light wave was demonstrated in the 1980's \cite{Arimondo1979,Moskowitz83}. 

These seminal works led to a number of experimental and theoretical studies \cite{Bernhardt1981,Marte1992aa,Gould86,Durr1996,Durr1999,Wilkens1991,champenois2001}. In particular, two extreme regimes of diffraction are useful for discussions: the Raman-Nath and the Bragg regimes \cite{Keller1999,GUPTA2001}. The Raman-Nath regime corresponds to a short and intense atom-light interaction. In this regime, the scattering process leads to the population of many atom momentum states. On the other hand, a long interaction time and a shallow optical potential give rise to the Bragg regime \cite{Martin88} where only a single diffraction order is allowed by energy conservation. It provides efficient and adjustable two-port beam splitters which are required for implementing two-path interferometer geometries. These interferometers are of a great interest for accurate interferometric measurements as they allow a straightforward readout of the atomic phase shift.

Although most of the precision measurements in atom interferometry use 2-photon Raman beam splitters \cite{Dickerson2013,Gauguet2009,Fang_2016,Bidel:2018aa,Hu2013,Freier_2016,Rosi:2014aa,Morel:2020aa}, Bragg scattering has several attractive features for atom interferometry. First, in a Bragg interferometer, the atoms propagate in the same internal state which makes the atomic phase less sensitive to external fields. This property could be advantageous for accurate inertial sensors \cite{Altin13}. Interferometers with a single internal state are also important for accurate measurements of some atomic properties \cite{Decamps2020}, or geometrical phase shifts \cite{Gillot2014,Gillot2013}. Furthermore, the high-order Bragg diffraction allows an increase of the separation between the interferometer arms, and hence the interferometer sensitivity. The first atom interferometer based on Bragg diffraction was demonstrated by \citet{GiltnerPRL95} with a momentum separation up to six photon recoils ($6\hbar k$).

High-order diffraction in the Bragg regime \cite{GiltnerPRA95, Koolen02} requires a very long atom-optical lattice interaction time which is difficult to handle in interferometric measurements. Nevertheless, in between the Raman-Nath and the Bragg regime, the so-called quasi-Bragg regime, it is possible to achieve an effective two-port beam splitter using a smooth temporal profile of the optical lattice intensity \cite{Keller1999,JansenPRA07,MullerPRA08}. Using a Gaussian shape pulse, \citet{MullerPRL08} performed the atom interferometer with the highest diffraction order ($24\hbar k$). A convenient description of the quasi-Bragg regime relies on the adiabatic following of eigenstates \cite{Keller1999,JansenPRA07,Gochnauer2019,SiemszPRA20}. Taking a full advantage of the adiabatic theorem \citet{SiemszPRA20} have demonstrated an analytical formalism describing losses and phase shifts for a sharp momentum distribution. 

Many efforts have been made to further increase the number of photon recoils transferred to the atom, in order to produce Large Momentum Transfer beam splitters (LMT). Different processes were demonstrated, most of them create a superposition of two momentum states using a first diffraction pulse followed by an acceleration of each state with a sequence of diffraction pulses \cite{Gupta2002, Chiow2011,Plotkin2018,Rudolph2020} or Bloch oscillations in an accelerated optical lattice \cite{Clade2009, Muller2009, Debs2011, McDonald2013}. Recent years have seen spectacular breakthroughs in the development of LMT-interferometers, demonstrating coherence for up to 408$\hbar k$ \cite{Gebbe2021}. 
The LMT beam splitters are one of the main prospects for improvement of inertial sensors technologies \cite{Hensel:2021aa,Li2021} and fundamental constant measurements \cite{Parker2018, Plotkin2018}. The LMT beam splitters are also central for new tests in gravitational physics \cite{Dimopoulos2007,Tino2021,Canuel20,graham2017,Zhan20,El-Neaj2020}. In addition, interferometers with a very large spatial separation \cite{Garcia06,Kovachy15} pave the way for measurements where the macroscopic separation between the two arms is essential, for example in cavity QED \cite{Durr1998aa}, entangled state engineering \cite{Islam19} or measurements such as Aharonov-Bohm related phases \cite{champenois01,Arvanitaki2008,Hohensee2012}. 

A thorough understanding of the quasi-Bragg scattering is fundamental for metrology with LMT atom interferometers. In particular, the  complex dynamics of the non resonant momentum states leads to diffraction phase shifts \cite{Buchner2003, MullerPRA08, SiemszPRA20} and multiple interferometers \cite{Altin13, Parker2016} which are crucial for determining the fundamental limits of the sensitivity and the accuracy. Theoretical approaches based on perturbative expansion allow a qualitative description of the diffraction phases in the quasi-Bragg regime \cite{MullerPRA08}. However, a more quantitative description needs to consider the momentum distribution of the atomic ensemble  \cite{SzigetiNJP12,SiemszPRA20}.

In this paper, we investigate experimentally the interplay between diffraction efficiency, diffraction phase and multiport beam splitters which have been explored theoretically in \cite{MullerPRA08,SzigetiNJP12,SiemszPRA20}. We measure the Rabi oscillations between the Bragg states and the non-adiabatic losses for high-order diffraction at finite momentum width. The very good agreement with numerical simulations allows a quantitative understanding of the diffraction dynamics and the diffraction phases. We demonstrate a Short (SP) and a Long Pulse (LP) regime within the quasi-Bragg regime. In the SP regime a complex dynamics is induced by large non-adiabatic losses, while in the LP regime the evolution merges the effective two-level model at a high Rabi cycle ($\ge \pi$). An optimal efficiency is found at the border between these two regimes.  We also present the interference patterns for multiport interferometers operating in the quasi-Bragg regime. The fringe patterns of each output ports of the interferometer are compared with numerical simulations. It allows to anticipate further experimental and theoretical studies regarding the inherent diffraction phase-shifts and multi-port features of LMT interferometers based on quasi-Bragg diffraction.

The paper is organized as follows: section \ref{model} presents general aspects of atom diffraction in the quasi-Bragg regime, and the numerical model we use to analyse the experimental results. Section \ref{exp} discusses the experimental diffraction patterns and the interferometers signals in the quasi-Bragg regime.

%%%%%%%%%%%%%%%%%%%%%%%%%%%%%%%
%%%%%%%%%%%%%%%%%%%%%%%%%%%%%%%
\section{Atom diffraction model}
\label{model}
%This section briefly review theoretical aspects of atomic diffraction for the discussion of the experimental results presented later. 
We consider a BEC interacting with a vertical optical lattice created by two counter-propagating laser beams with a frequency difference $\Delta \omega = \omega_1 - \omega_2$. The mean wave-number of the two running waves is denoted $k = \frac{\omega_1+\omega_2}{2 c}$. The laser is far detuned from the resonant absorption frequencies allowing an adiabatic elimination of the excited internal state. Therefore the atom-light interaction reduces to a light shift, proportional to the light intensity, leading to the potential $2\hbar \Omega \sin^2{(kz)}$, where $\Omega$ is the two-photon Rabi frequency (see appendix \ref{ap1}). 
\begin{figure}
\centerline{\includegraphics[width=0.4\textwidth]{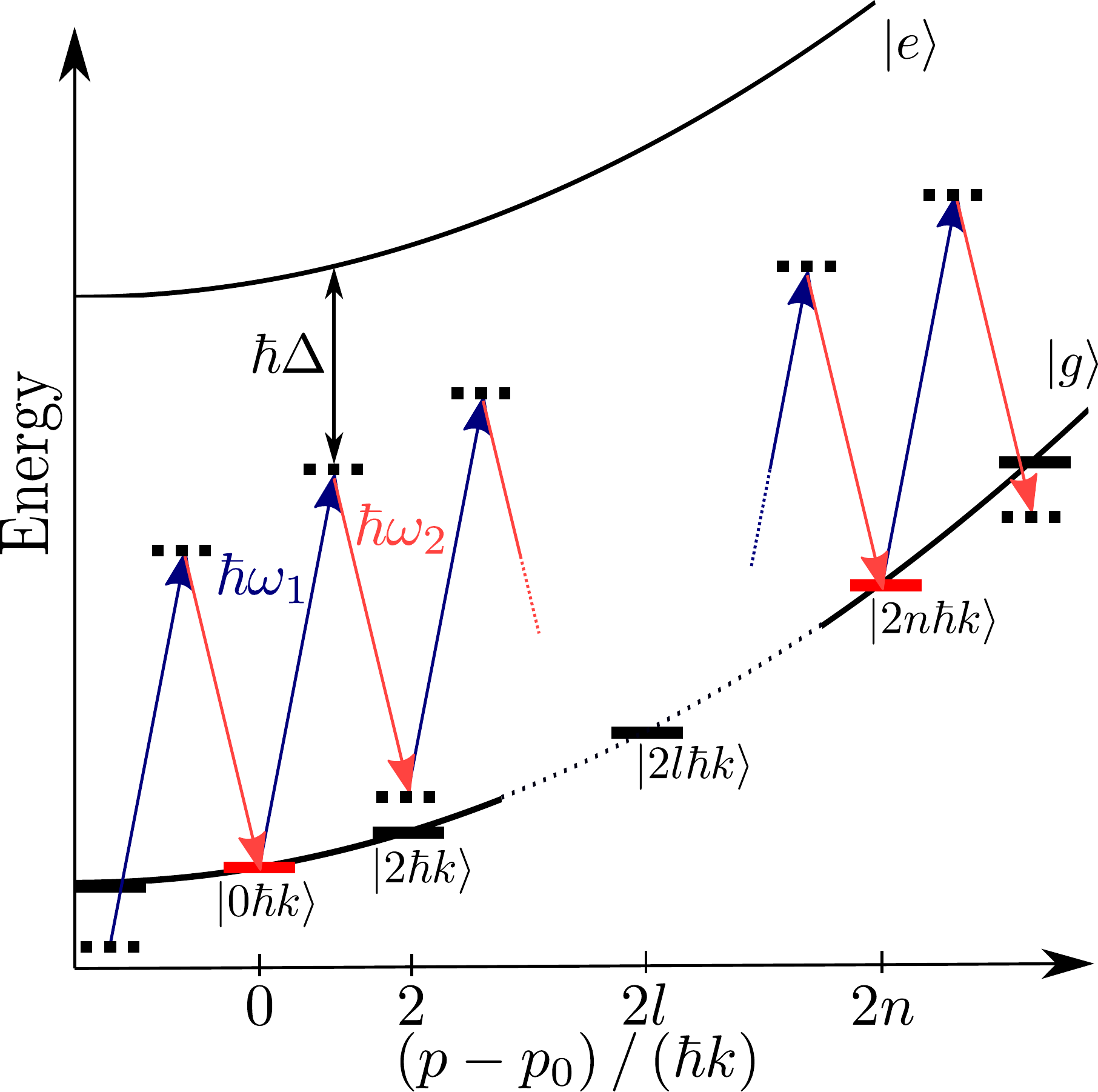}}
\caption{(colors online) Energy-momentum parabolic dispersion relation. The Bragg order $n$ diffraction process can be interpreted as multiple two-photon processes between an input state in momentum $\ket{p_0}$ and an output momentum state $\ket{p_0+2n\hbar k}$. The two laser beams forming the optical lattice have frequencies $\omega_1$ and $\omega_2$ detuned by $\Delta$ with respect to the atomic transition. }
\label{parabol}
\end{figure}

In our experiments, we use an expanded BEC source where atom-atom interactions are negligible. Therefore, the evolution of the BEC is completely described by a single atom Hamiltonian $H$ including the kinetic energy, the gravitational potential and the periodic light shift. In the laboratory frame, the Hamiltonian writes:
\begin{equation}
 H = \frac{\hat p^2}{2M} + M g \hat z + 2\hbar \Omega(t) \sin^2\bigg[ k \hat z -\frac{\phi (t)}{2} \bigg]
\end{equation}
Where $M$ is the mass of the atom, $\hat{z}$ and $\hat{p}$ the position and momentum operators along the direction of the optical lattice. $\phi(t)$ is the phase difference between the two laser running waves of the optical lattice. Using the unitary operator $ U = e^{\frac{i}{\hbar}\frac{M g^2 t^3}{6}} e^{\frac{i}{\hbar}\frac{\phi(t)}{2k}\hat p} e^{\frac{i}{\hbar}M g t\hat z}$, we transform the Hamiltonian $\tilde H =  U  H  U^{\dagger} + i \hbar \dot{U} U^{\dagger} $ corresponding to a free falling frame and, in this Hamiltonian, the gravitational potential does not appear explicitly:
 \begin{equation}
    \tilde H   =  \frac{\hat p^2}{2M} - \hat p \big(\frac{\dot{\phi}}{2 k} +gt\big) + \hbar \Omega(t) - \frac{\hbar \Omega(t)}{2} (e^{2ik\hat{z}}+e^{-2ik\hat{z}})
\label{tildH}
 \end{equation}
The operators $e^{\pm 2ik\hat{z}}$ couple atomic states differing by two photon momenta. Therefore, the periodic potential is interpreted as a two-photon process where a photon is absorbed from one running  wave and re-emitted into the other one with a transfer of two photon momenta $2\hbar k$. The two-photon process can occur $n$ times and transfer $2n$ photon momenta corresponding to higher diffraction orders as illustrated in the figure \ref{parabol}. Hence, it is convenient, to expand the wave-function on the plane waves $\ket{2l\hbar k}$: $\ket{\Psi} = \sum_{l=-m}^{n+m} A_{l} \ket{2l\hbar k}$ where $A_l$ are the complex amplitudes of the plane wave decomposition (see fig.\ref{parabol}). We truncate the basis considering $2m$ outer states surrounding the two Bragg states $\ket{0 \hbar k}$ and $\ket{2n\hbar k}$ and all the states in between them. In this basis, the Hamiltonian $\tilde H(t)$ is written as a tridiagonal matrix:
\begin{equation}
\tilde H(t) = 4 \hbar \omega_r
\begin{pmatrix}
\delta_{-m} & \gamma(t) & 0 & \dots &\dots & \dots & 0&  \\
\gamma(t)^* &  \ddots & \ddots & \ddots & \ddots &\ddots &\vdots \\
 0 & \ddots  & 0  &  \ddots & \ddots &\ddots &\vdots \\
\vdots & \ddots & \ddots & \ddots & \ddots & \ddots & \vdots \\
\vdots & \ddots & \ddots & \ddots &  \delta_n &\ddots & 0 \\
\vdots & \ddots & \ddots & \ddots & \ddots & \ddots & \gamma(t) \\
0 & \dots & \dots & \dots & 0 & \gamma(t)^* & \delta_{n+m} \\
\end{pmatrix},
\label{H}
\end{equation}
where $\omega_r = \frac{\hbar k^2}{2 M}$ is the recoil frequency of a single photon and $\gamma(t)=\frac{\Omega(t)}{8 \omega_r}$ the dimensionless effective two-photon Rabi frequency $\Omega(t)$. The diagonal terms $\delta_l (t) = l^2 - l \tilde v(t)$ depend on the usual kinetic energy in $l^2$ and on the velocity $\tilde v(t)$ of the lattice with respect to the free falling atoms (in units of $v_r=\frac{\hbar k}{M}$): 
\begin{equation}
    \tilde v(t) = \frac{v(t)+(gt-v_0)}{v_r} = \frac{\Delta \omega(t)}{4 \omega_r}+\frac{\left(gt-v_0\right)}{v_r}.
\label{vtilde}
\end{equation}
$v_0$ is the  initial atom velocity in the lattice direction in the laboratory frame. In addition, the lattice velocity is set by the frequency difference $\Delta\omega$ in between the two beams. A time dependent frequency ramp $\Delta \omega(t)=\omega_0+\alpha t$, with $\alpha t=-2kgt$, is adjusted to cancel out the Earth acceleration so that the Bragg condition is always verified. The constant value $\omega_0=4n\omega_r+2kv_0$ cancels out the diagonal term $\delta_n$ and defines the Bragg condition at the order $n$ for a given $v_0$. 

The Schr\"odinger equation leads to a system of differential equations. Approximate solutions exist, in particular for rectangular pulse shapes of the lattice amplitude. Two extreme cases are widely discussed in the literature \cite{champenois2001}. The first case corresponds to short enough interaction times to neglect the dynamics of the external states, \textit{i.e.} for short interaction times compared to the period of classical oscillation in the optical potential $\tau < \left(8 \omega_r \sqrt{\left|\gamma \right|}\right)^{-1}$. This approximation corresponds to the Raman-Nath approximation \cite{Raman36}, introduced for the diffraction of light by high frequency sound waves, now used in acousto-optic devices \cite{klein1967,Gaylord81}. The broad spectrum associated with a short rectangular pulse leads to multiple diffraction orders which is poorly suited for atom beam splitters. On the other hand, the Bragg regime corresponds to the interaction with a weak potential: $ \left|\gamma \right| \ll 1$. The population in the non resonant momentum states vanishes, and they can be ignored in a perturbative approximation. The result is a two-level system leading to Rabi oscillations between the diffracted states. 

A rough estimate of the effective Rabi frequency for diffraction of order $n$ is given by the first non-zero term of the perturbation expansion \cite{GiltnerPRA95}: $\Omega_{n} \approx \frac{\Omega^n}{(8\omega_r)^{n-1}(n-1)!^2}$.
This result is not accurate as soon as $\Omega_n$ is not very small with respect to $\omega_r$, the higher terms of the perturbation expansion are then no more negligible. However, this quantity is proportional to the difference of the $a_n$ and $b_n$ coefficients of the Mathieu equation \cite{abramowitz64}, and it is well established that the power expansions of these coefficients have a rather small radius of convergence \cite{Meixner1980}, so that the convergence of the perturbation expansion of $\Omega_n$ is slow. Despite its inaccuracy the approximate formula for $\Omega_{n}$ illustrates the necessity of high laser power and a very long interaction for large values of the Bragg order $n$. Therefore, the Bragg regime is not of practical interest for high momentum transfer.

We are interested in the intermediate regime introduced by \citet{MullerPRA08} and named the quasi-Bragg regime. This regime has shown efficient high-order Bragg diffraction for an experimentally accessible set of parameters. The regime occurs when the potential is switched on and off smoothly. As for the Bragg diffraction, in the quasi-Bragg regime two output momentum states are mainly populated once the pulse has ended. In the analytical theory developed in \cite{SiemszPRA20} the finite momentum dispersion is described perturbatively and the losses act as small corrections to the description of the quasi-Bragg diffraction process. As the parameters of our experiments are beyond these approximations, we study the dynamics by numerically solving the Schr\"odinger equation in the free falling frame using the discrete momentum basis. The populations on each momentum states are calculated as the squared complex amplitudes after the interaction with the lattice: $P_l=A_l^2$. In what follows, we are using a Gaussian pulse of the lattice amplitude to reduce losses in unwanted states \cite{MullerPRA08,SzigetiNJP12}:
\begin{eqnarray}\label{EqGauss}
\gamma(t)=\gamma_{\mathrm{max}}  \exp \left[- \frac{t^2}{2\sigma^2}\right],
\end{eqnarray}
where $\sigma$ is the Gaussian pulse duration and $\gamma_{\mathrm{max}}$ the amplitude (peak two-photon Rabi frequency). The numerical propagation is performed on a $10 \sigma$ time window centered on the pulse.\\

\begin{figure}
\centerline{\includegraphics[width=0.5\textwidth]{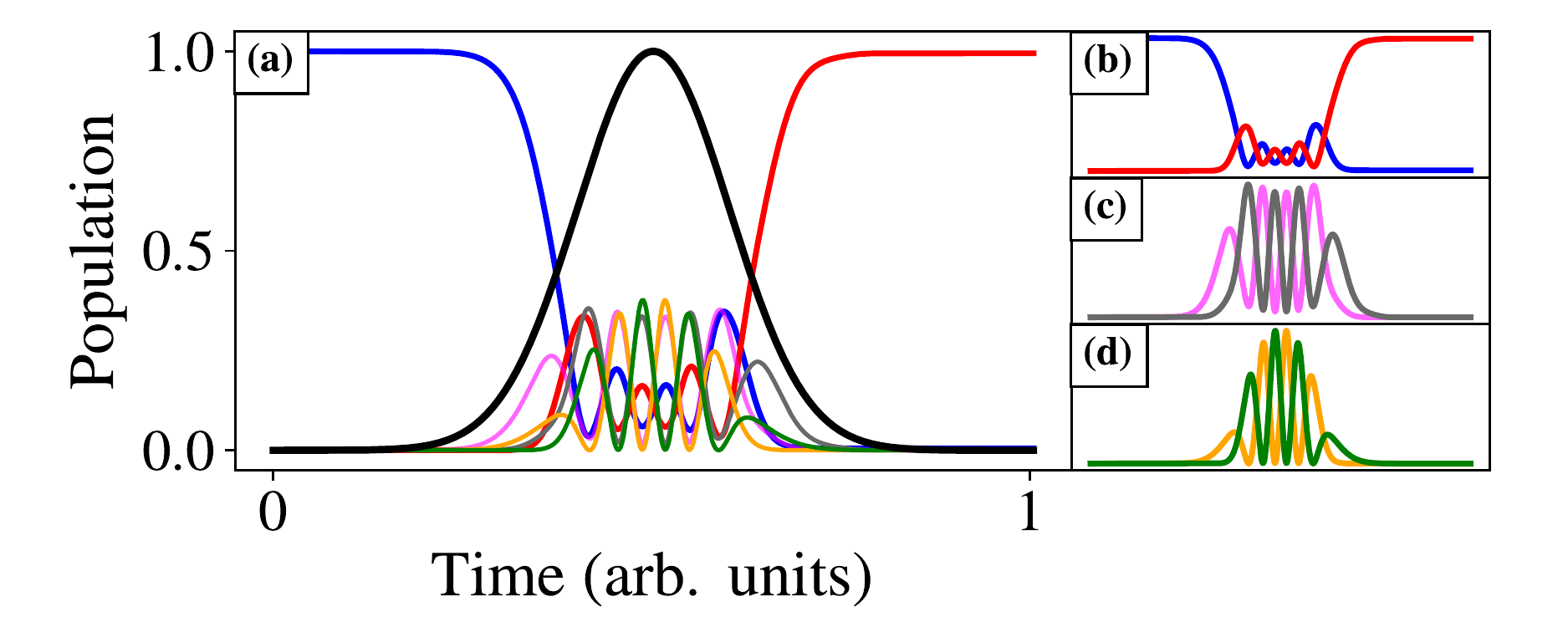}}
\caption{(colors online) (a) Evolution of the populations during a Gaussian pulse for Bragg order $n=3$. The solid black line represents the Gaussian pulse amplitude as a function of time. The pulse duration $ \sigma = \omega_r^{-1}$ and the maximum amplitude $\gamma_ {\mathrm{max}} = 3.3$ have been chosen to produce a perfect Bragg mirror. Populations are plotted in (b) for the two Bragg states ($\ket{0\hbar k}$ in blue and $\ket{6\hbar k}$ in red) and in (c) and (d) for the unwanted states. Inner states ($\ket{2\hbar k}$ in pink and $\ket{4\hbar k}$ in grey) (c) and outer states ($\ket{-2\hbar k}$ in yellow and $\ket{8\hbar k}$ in green) (d) oscillate by pair.}
\label{gauss}
\end{figure} 
Figure \ref{gauss} illustrates the time evolution of the populations in the momentum states during a Gaussian pulse of Bragg order $n=3$ and for zero velocity dispersion. The pulse parameters are set to realise a perfect mirror pulse where only the Bragg state $\ket{6\hbar k}$ is populated at the end of the pulse.

If the Bragg condition is fulfilled, the two Bragg states are two eigenstates and are degenerate at the start of the interaction ($\left|\gamma \right|  \ll 1$). When the amplitude of the lattice increases sufficiently slowly, the populations adiabatically follow these two eigenstates. At the end of the pulse, the two eigenstates return back to the initial two momentum states leading to Rabi oscillations. The weight in each Bragg state depends on the accumulated phase (Rabi phase $\theta_\mathrm{R}$) during the evolution. For the calculation, we use the momentum basis which is not the eigenbasis of the system (Bloch states) in the presence of the lattice. The adiabatic following of the two eigenstates corresponds to the oscillation of unwanted momentum states by pairs with a non-negligible amplitude which finally destructively interfere.

The complex dynamics between momentum states during the pulse is a signature of the quasi-Bragg regime. It emphasizes the importance of considering the inner states ($0<l<n$) and some outer states ($l<0$ and $l>n$) to accurately describe the complete evolution of the atomic populations. The number $2m$ of outer states to be considered is linked to the spectral width of the pulse and so depends on the Bragg order $n$, the pulse duration $\sigma$ and on the pulse amplitude $\gamma_{\mathrm{max}}$. The appendix~\ref{ap3} discusses a criterion on the minimum number of outer states to be taken into account in~\eqref{H} to ensure the convergence of our simulations.
\begin{figure}
\centerline{\includegraphics[width=0.5\textwidth]{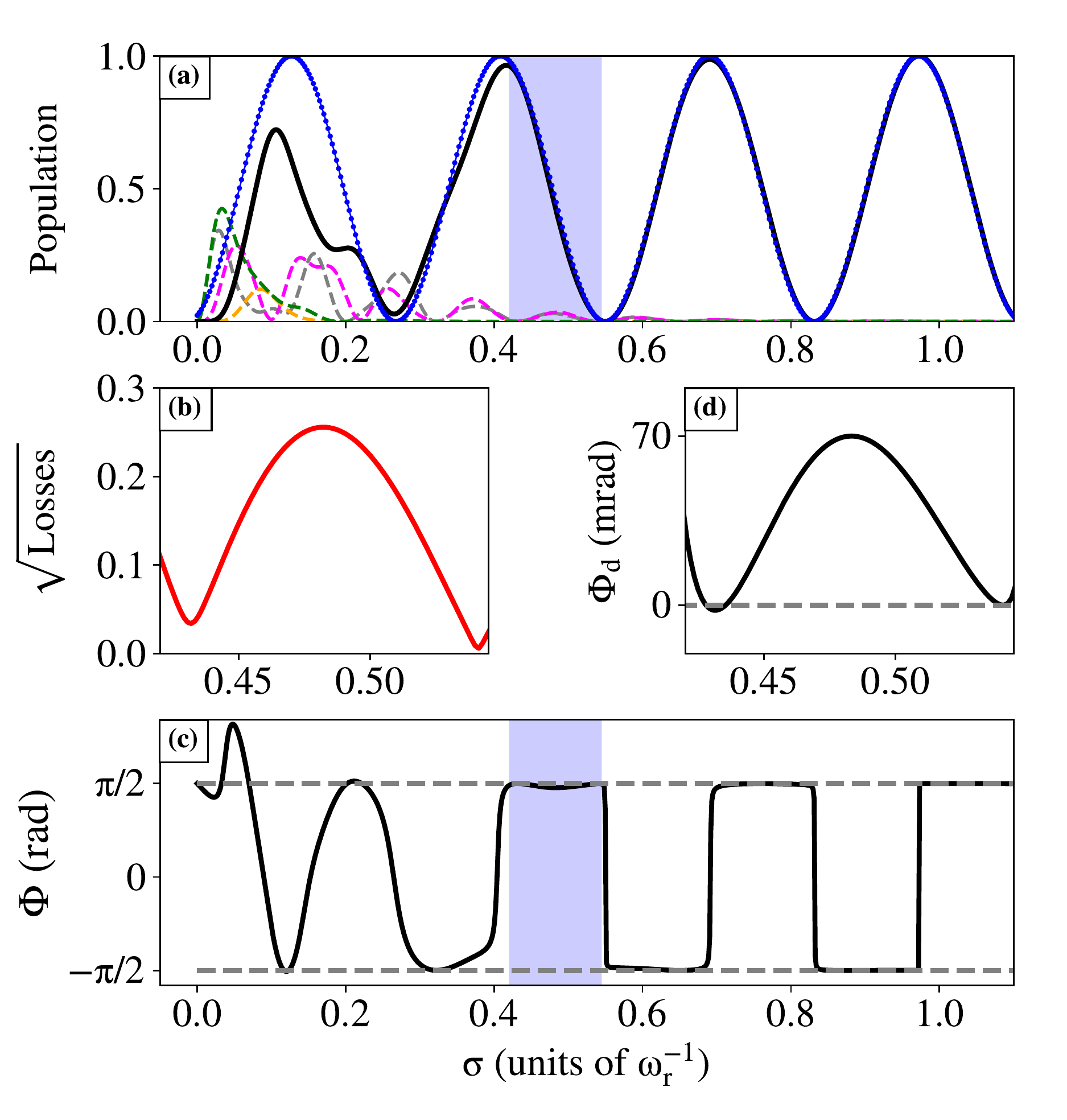}}
\caption{(colors online) Numerical simulation of diffraction at order $n=3$ in the quasi-Bragg regime with zero velocity dispersion and $\gamma_{\mathrm{max}}=3.3$. (a) Evolution of populations in the different momentum states as a function of the pulse duration $\sigma$, the population of the initial state $|0 \hbar k \rangle$ is not plotted. The black solid line represents the population in the diffracted Bragg state $\ket{6\hbar k}$; the populations in the unwanted states are plotted in dashed lines. The blue dotted line is the result of an effective two-level calculation for the population in the $|6\hbar k\rangle$ state. It overlaps with the numerical calculation in the Long Pulse (LP) regime and shows that the Rabi phase $\theta_\mathrm{R}$ is continuously built from the Short Pulse (SP) regime. (b) Square root of the total population in the unwanted states for pulse duration around $0.5\omega_r^{-1}$ (shaded region in (a) and (c)). (c) Diffraction phase $\Phi$ as a function of $\sigma$. The two gray dashed lines indicate constant $\pm\pi/2$ phases. They match with the diffraction phase calculated in LP regime. (d) Diffraction phase compensated from the trivial $\pi/2$ contribution ($\Phi_\mathrm{d}=\Phi - \pi/2$) for the same pulse duration.}
\label{rabi}
\end{figure}

In the quasi-Bragg regime, changing the pulse duration for a given amplitude does not only drive Rabi oscillations between the two Bragg states. Figure~\ref{rabi}.a shows the evolution of the populations in the different relevant states at the end of a pulse of duration $\sigma$ for $\gamma_{\mathrm{max}}=3.3$. Two sub-regimes can be distinguished: the Short Pulse (SP) regime, where many unwanted states are populated and the Long Pulse (LP) regime, where the population is oscillating in between the two Bragg states. In the LP regime, the oscillation between these two states is well predicted by a two-level calculation despite the fact that unwanted states are transiently populated during the pulse. It is interesting to notice that in the SP regime the populations in the unwanted states, \textit{i.e.} the losses, synchronously cancel out for specific values of $\sigma$. It offers the opportunity to experimentally choose pulse parameters that exploit the loss cancellation in the SP regime to maximise the diffraction efficiency. In what follows, the diffraction efficiency refers to a mirror pulse, and it is measured as the fraction of the population transferred in $\ket{2n \hbar k}$, i. e. $P_n$.

\begin{figure}
\centerline{\includegraphics[width=0.45\textwidth]{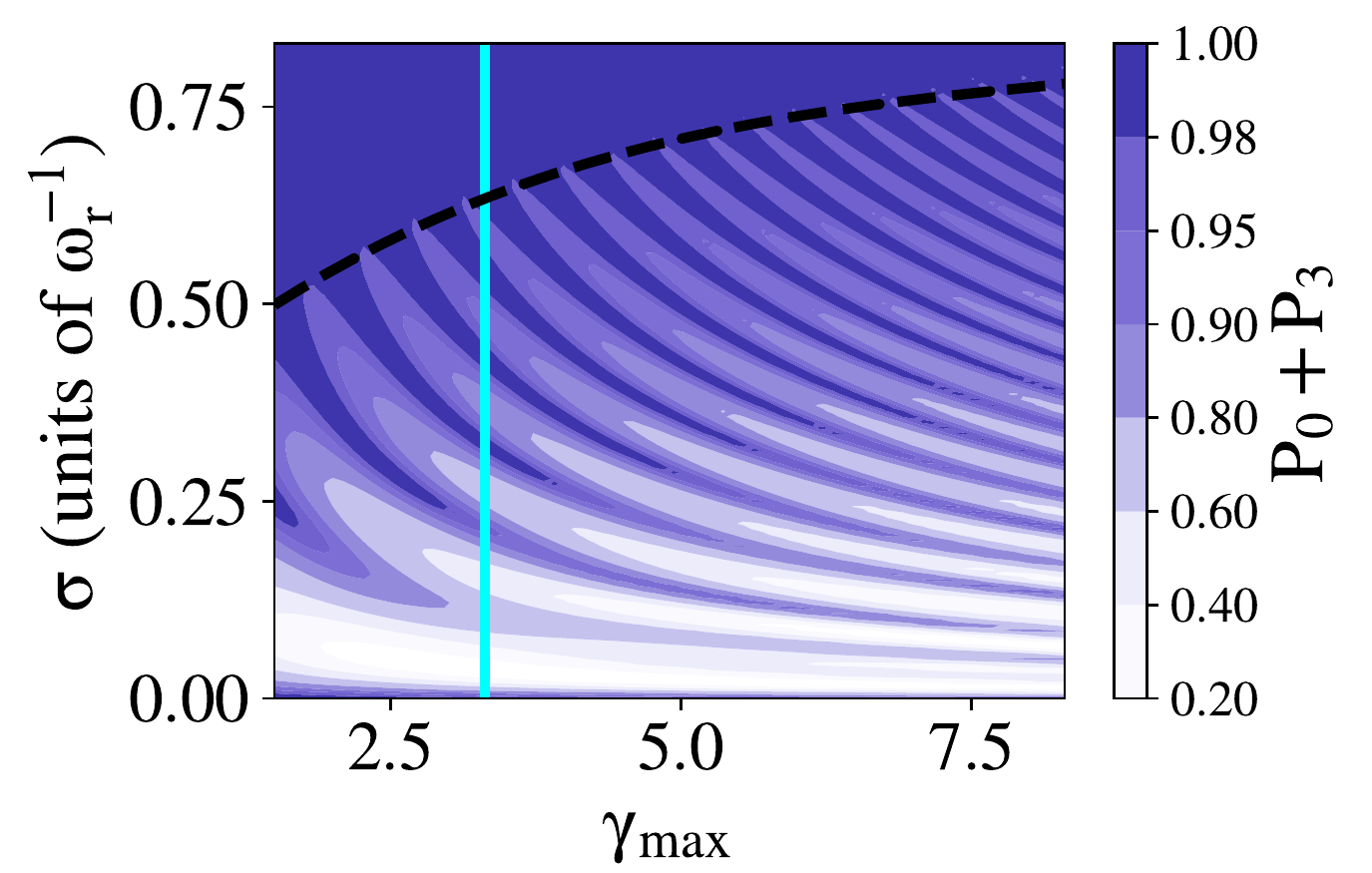}}
\caption{(colors online) The total population $P_0 + P_3$ of the two Bragg states ($\ket{0 \hbar k}$ and $\ket{6 \hbar k}$) in color scale in a $\sigma-\gamma_{\mathrm{max}}$ map for $n = 3$. The calculation assumes a vanishing velocity dispersion. The black dashed line is a guide to the eye showing the boundary between the SP and LP regimes. The light blue line corresponds to the cut at $\gamma_{\mathrm{max}}= 3.3$ studied in Figure \ref{rabi}.}
\label{dents}
\end{figure}

The boundary between the SP regime and the LP regime as well as the periodic cancellation of the losses are illustrated in fig.~\ref{dents}. It presents the total population in the two Bragg states in a $\sigma-\gamma_{\mathrm{max}}$ map for the Bragg order $n=3$. 
The SP and LP regimes are clearly identified. The boundary is pushed towards longer pulses as $\gamma_{\mathrm{max}}$ increases. It follows the intuitive picture that the spectral width increases with $\gamma_{\mathrm{max}}$ and needs to be compensated by a longer pulse to maintain the adiabaticity of the pulse. The actual shape of the boundary is however non-trivial since it relies on multiple interferences between unwanted states. It also appears that the losses cancellation presents a periodic pattern that can be interpreted as St\"uckelberg interferometers \cite{SHEVCHENKO2010}. The St\"uckelberg interferences are due to the symmetry of the Gaussian pulse, which induces a symmetric series of avoided crossings between the rising part and the falling part of the pulse. Each avoided crossing acts as beam splitters for the eigenstates, which splits and recombines the wavefunctions. The resulting interferometers lead to oscillations between the Bragg states and the losses that depends on the phase difference due to the propagation in the different eigenstates. Following \cite{SiemszPRA20}, we call these losses due to non adiabatic effect the Landau-Zener (LZ) losses.

The diffraction efficiency is not the only relevant parameter when planning to use Bragg diffraction to build an atom interferometer. The diffracted waves are imprinted with a phase that needs to be controlled at a metrological level. The numerical calculation gives access to the diffraction phase $\Phi$ between the two Bragg states as the difference between the arguments of the complex amplitudes after the pulse: $\Phi=\arg{A_n}-\arg{A_0}$.

Figure~\ref{rabi}.c represents the relative phase $\Phi$ as a function of $\sigma$ for $\gamma_{\mathrm{max}}=3.3$. In the LP regime, the diffraction phase oscillates as expected in the deep Bragg regime (effective two-level approximation), \textit{i.e} the phase $\Phi$ jumps between $\pm \pi/2$ for each half period. However, in the SP regime, we observe a complex phase evolution highlighting the link between diffraction phase and LZ losses \cite{MullerPRA08,SiemszPRA20}. \citet{MullerPRA08} have shown that the non-trivial phase $\Phi_\mathrm{d}$ is bounded by $\Phi_\mathrm{d} \leq \sqrt{\mathrm{Losses}}$. Figures~\ref{rabi}.b and d present respectively the square root of the losses and $\Phi_\mathrm{d}$ for $\sigma\in[0.37\omega_r^{-1},\ 0.52\omega_r^{-1}]$ close to the LP regime boundary where the losses are small. In this range, $\Phi_\mathrm{d} \approx  \sqrt{\mathrm{Losses}}/3$, it confirms that the vanishing losses coincide with a minimum of non-trivial diffraction phase.

To reproduce experimental data as accurately as possible, the finite velocity dispersion of the atomic cloud has to be taken into account in the calculation as the diagonal matrix elements explicitly depends on the atom velocity. It is done by adding to the lattice velocity $\tilde{v}(t)$ used in eq.~\ref{vtilde} a random variable $\delta \tilde{v}$ following the Gaussian distribution:
\begin{equation}
    \rho(\delta \tilde{v})=\frac{1}{\sqrt{2\pi\sigma_{\tilde{v}}^2}}\exp[-\frac{{\delta \tilde{v}}^2}{2\sigma_{\tilde{v}}^2}],
\end{equation}
where $\sigma_{\tilde{v}}$ is a Gaussian width proportional to the atomic cloud velocity dispersion. The numerical calculation is performed for each velocity class and the mean atomic population in each momentum state is the weighted average over the distribution. \\ 

\begin{figure}
\centerline{\includegraphics[width=0.5\textwidth]{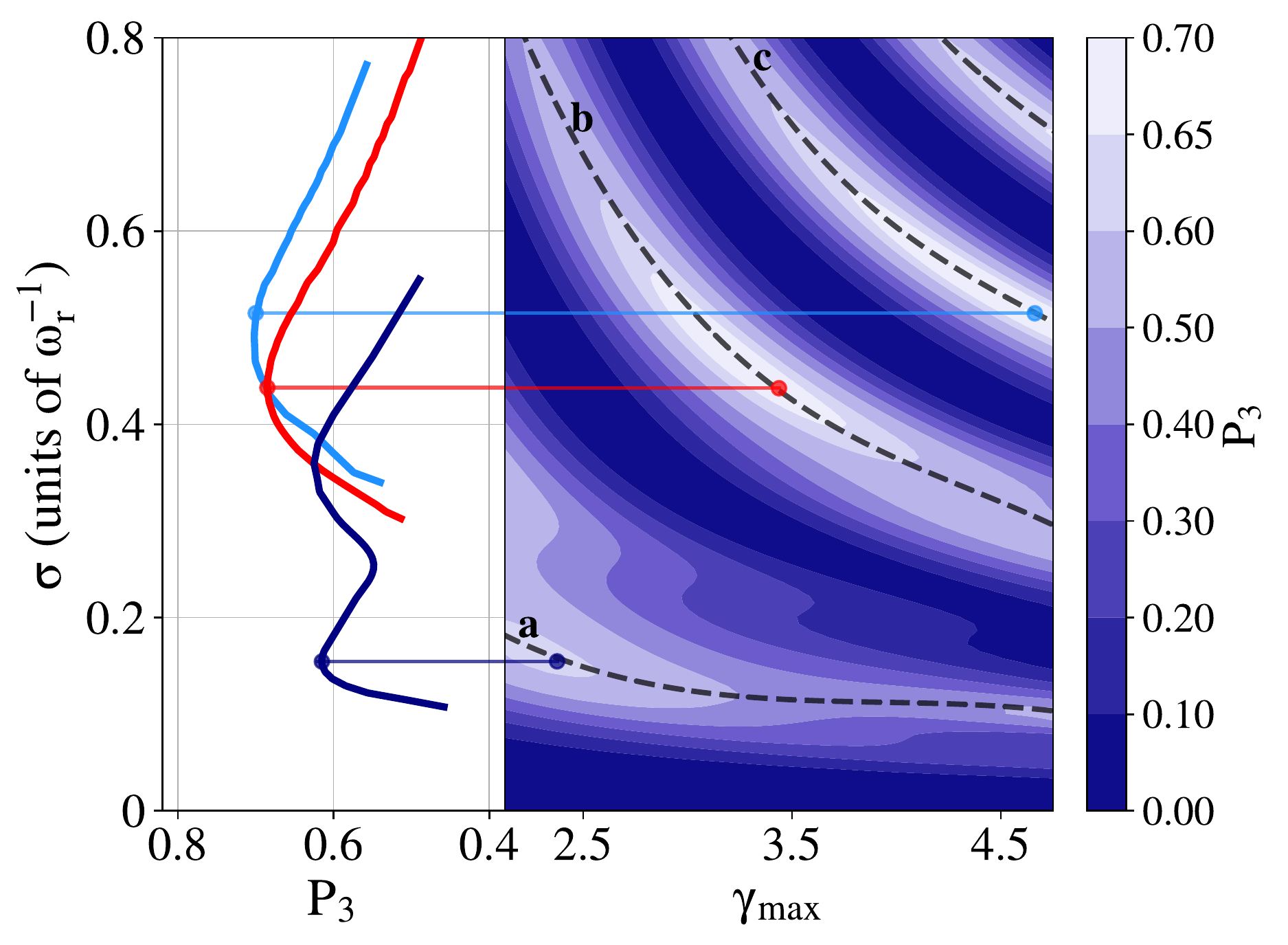}}
\caption{(colors online) Population $P_3$ in the diffracted Bragg state for $n=3$. The calculation takes into account a velocity dispersion of $0.32 v_r$. (right) Diffracted population $P_3$ in color scale in a $\sigma - \gamma_{\mathrm{max}}$ map. Dashed lines are guides to the eye showing local maxima lines achieving Rabi phases $\theta_{\mathrm{R}}= \pi$ (dashed line $a$), $3\pi$ (dashed line $b$), $5\pi$ (dashed line $c$), etc. (left)  Diffracted population along maxima lines. The dark blue curve corresponds to $\theta_{\mathrm{R}}=\pi$, the red curve corresponds to $\theta_{\mathrm{R}}=3\pi$, and the light blue curve corresponds to $\theta_{\mathrm{R}}=5\pi$. An optimal transfer is realised for well chosen $(\sigma, \gamma_{\mathrm{max}})_{\rm{opt}}$ couples for $\theta_{\mathrm{R}}\ge 3\pi$.}
\label{sigop}
\end{figure}

The main consequence of a finite velocity dispersion is to reduce the diffraction efficiency for long pulses as part of the atomic distribution ends up being off resonance. From this point of view, one would like to make a compromise on the pulse duration to minimize both Landau-Zener losses and velocity selection losses \cite{SzigetiNJP12}. The existence of this compromise is illustrated in fig.~\ref{sigop} taking as an example the diffraction order $n=3$ with a velocity dispersion of $0.32v_r$ relevant to our experiment (see section~\ref{exp}). The right part of fig.~\ref{sigop} maps the population in the diffracted Bragg state $\ket{6 \hbar k}$ in a $\sigma - \gamma_{\mathrm{max}}$ plane. It presents a series of local maxima corresponding to Rabi phases that match odd multiples of $\pi$. The left part of fig.~\ref{sigop} shows the evolution of the transferred population as a function of the pulse duration along the local maximum lines drawn in dashed lines in the $\sigma - \gamma_{\mathrm{max}}$ plane. Each curve corresponds to a given Rabi phase ($\theta_\mathrm{R}=\pi$, $3\pi$, etc.). All the odd-$\pi$ Rabi phases in the LP regime reach a similar maximum transfer efficiency for a given $(\sigma,\gamma_{\mathrm{max}})_{\rm{opt}}$. This optimal transfer is limited by the velocity selection losses which are similar for all $(\sigma,\gamma_{\mathrm{max}})_{\rm{opt}}$. 
In this example, the maximum efficiency can be reached for $\theta_{\mathrm{R}}\ge 3\pi$. An interesting tradeoff is the pulse duration $\sigma_\mathrm{opt} \simeq 0.4\omega_r^{-1}$ corresponding to $\theta_{\mathrm{R}}=3\pi$, which minimises the peak Rabi frequency. Going for higher Rabi phase with equivalent efficiency would further reduce Landau-Zener losses but it would increase negative effects scaling with $\sigma \gamma_{\mathrm{max}}$ such as spontaneous emission losses or AC-Stark shifts related phases.

In the next section we numerically and experimentally investigate the impact of finite temperature on high-order diffraction beam splitters in the quasi-Bragg regime. 

%%%%%%%%%%%%%%%%%%%%%%%%%%%%%%%
%%%%%%%%%%%%%%%%%%%%%%%%%%%%%%%

\section{Experimental results}
\label{exp}
\subsection{Experimental setup}
The atomic source is an evaporatively cooled Rubidium 87 ensemble in a crossed dipole trap in the presence of a magnetic field gradient to purify the ensemble in the magnetically insensitive Zeeman sublevel of the lower hyperfine state $\ket{F=1,m_F=0}$. We obtain a Bose-Einstein Condensate (BEC) of $N=8\times 10^4$ atoms with final trapping frequencies $(60\times900\times1100)~\mathrm{Hz}^3$. The single-shot velocity dispersion is further reduced by transferring the BEC in a much shallower trap $(10\times80\times80)~\mathrm{Hz}^3$, leading to a velocity dispersion of about $0.32\pm0.05v_r = 1.9\pm0.3~\mathrm{mm.s}^{-1}$, with $v_r\sim5.9~\mathrm{mm.s}^{-1}$, including center-of-mass velocity fluctuations.

\begin{figure}
\centerline{\includegraphics[width=0.4\textwidth]{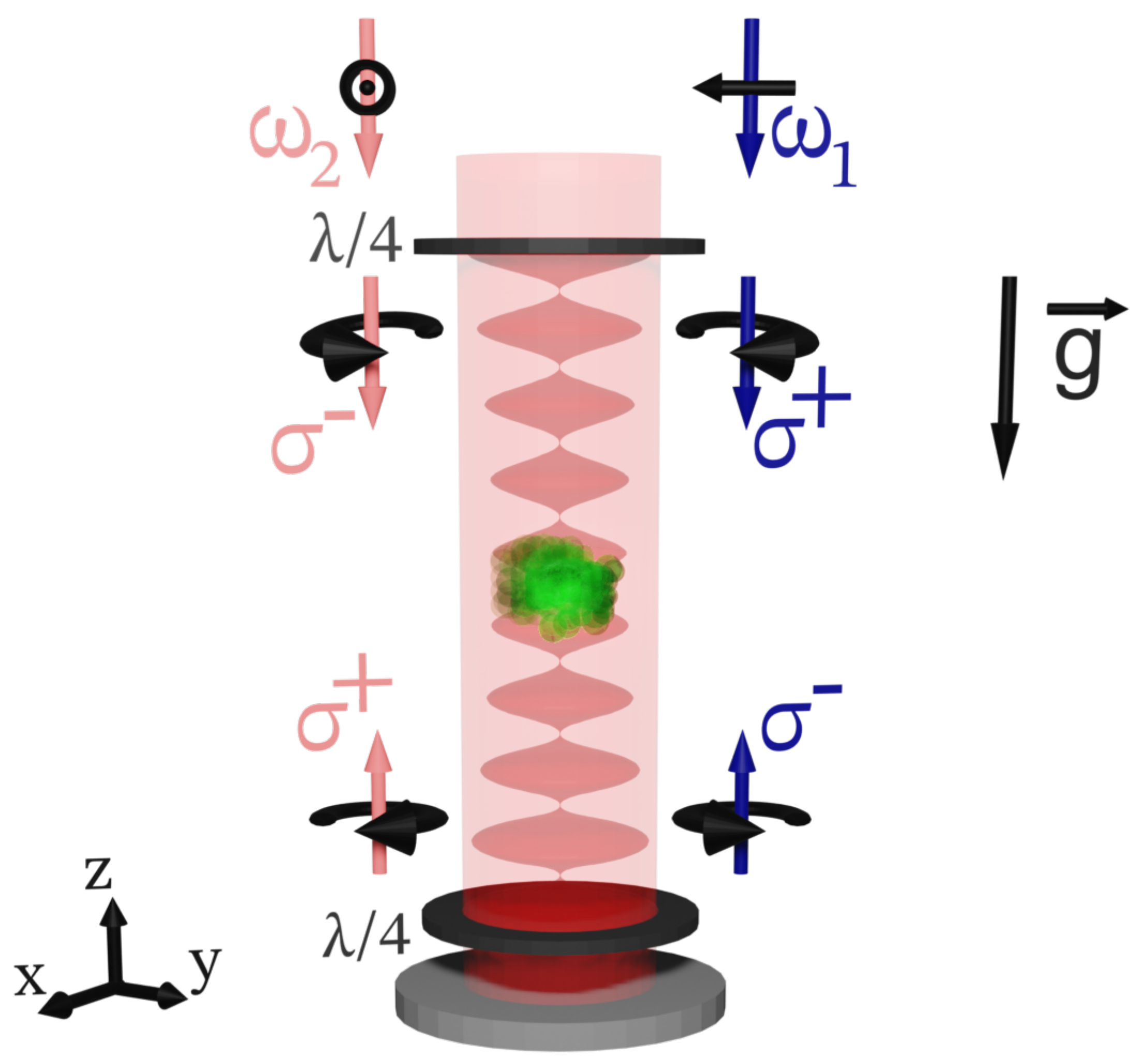}}
\caption{(colors online) Two laser beams at frequencies $\omega_1$ and $\omega_2$ are overlapped with orthogonal linear polarizations. The beams pass through a first $\lambda /4$ wave plate to obtain circular polarizations and are retro-reflected through a second $\lambda /4$. It creates two vertical optical lattices corresponding to the $\sigma^+/\sigma^+$ and $\sigma^-/\sigma^-$ pairs each pairing both frequencies.} 
\label{setup}
\end{figure}

The vertical optical lattice used for Bragg diffraction is sketched in fig.~\ref{setup}. The laser source is a $1560~\mathrm{nm}$-laser frequency doubled to $780~\mathrm{nm}$ with a detuning $\Delta=11~\mathrm{GHz}$ from the Rubidium 87 $\ket{5S_{1/2}, F=1}$ to $\ket{5P_{3/2}, F'=2}$ transition. This detuning is sufficient to ensure negligible spontaneous emission for the considered interaction times. 
However, for a given diffraction order, we have seen that there is an optimal $\gamma_{\mathrm{max}}$. Therefore, for a given laser intensity, it is possible to find an optimal $\Delta$ value, which minimizes the spontaneous emission \cite{SzigetiNJP12}.
The laser is split into two paths passing through two double-pass Acousto-Optic Modulator (AOM) controlling the frequency difference and phase of the optical lattice. They are recombined on crossed polarization in a polarization maintaining fiber. A last AOM controls the amplitude of both beams up to $150~\mathrm{mW}$ and so allows the temporal pulse-shaping of the optical lattice. The two beams are finally collimated with a Gaussian waist $w_0=4~\mathrm{mm}$ and polarized with orthogonal circular polarizations. The beams are then retro-reflected along the vertical axis after passing through a quarter waveplate. It creates two quasi-standing waves for each circular polarization with opposite velocities $\pm \Delta\omega/2k$ and opposite phases. 

For large enough initial velocity of the atoms $v_0 \gg v_r$, only one of the two lattices can fulfill the Bragg condition at a time. The other one is Doppler shifted off resonance and does not interact with the atoms. In practice, the initial velocity is acquired by letting the atom fall under gravity for $6~\mathrm{ms}$ before the optical lattice is turned on. To ensure the Bragg condition to be satisfied for long interaction times or complex pulse sequences, the constant gravity acceleration is compensated by a constant frequency ramp on one of the two beams (see eq.~\ref{vtilde}).

The populations in different momentum states are measured by time-of-flight fluorescence imaging after $20~\mathrm{ms}$ of free fall. The spatial separation corresponding to a momentum separation of $2 \hbar k$ equals around $230~\mu\mathrm{m}$ which is larger than the typical cloud size of $\simeq 60~\mu\mathrm{m}$ (radius at $\frac{1}{\sqrt{e}}$). The images are integrated along the direction orthogonal to the diffracting lattice and the resulting profiles are fitted with a sum of regularly separated Gaussians corresponding to each momentum state. The relative population in each state is taken as the integral of the corresponding Gaussian divided by the total integrated signal. 

%%%%%%%%%%%%%%%%%%%%%%%%%%%%%%%
%%%%%%%%%%%%%%%%%%%%%%%%%%%%%%%
\subsection{Quasi-Bragg diffraction}
\label{res}

\begin{figure*}
 \centerline{\includegraphics[width=0.9\textwidth]{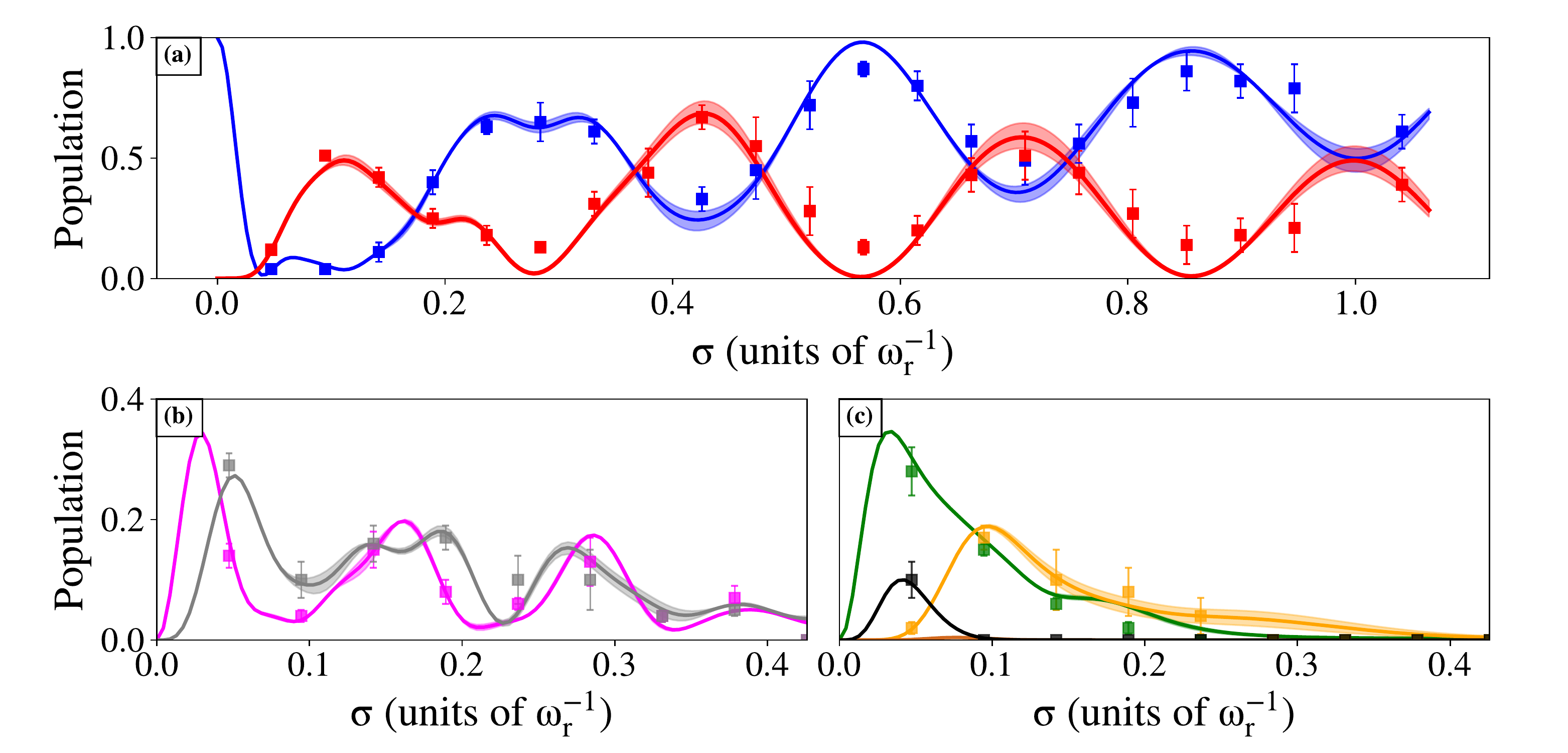}}
\caption{(colors online) Comparison of the measured and calculated relative populations of the various momentum states as a function of $\sigma$ for a quasi-Bragg diffraction of order $n = 3$. The peak two-photon Rabi frequency is $\gamma_{\mathrm{max}} = 3.3$. The relative populations in the different momentum states are plotted in panel a) blue and red for the two Bragg states $\ket{0\hbar k}$ and $\ket{6\hbar k}$ respectively. (b) Relative population in the inner states (pink: $\ket{2\hbar k}$ and grey: $\ket{4\hbar k}$). (c) Relative population in the outer states (black: $\ket{-4\hbar k}$, green: $\ket{-2\hbar k}$, and yellow: $\ket{8\hbar k}$). Each experimental point is an average of ten measurements and the error bar is the statistical shot-to-shot error. The solid lines are the results of the numerical simulation for a velocity dispersion corresponding to $0.32\pm0.05v_r$. The experimental uncertainty on the velocity dispersion is shown as the shaded thickness around the simulated populations.}
\label{figRabi}
\end{figure*}

The high-order Bragg diffraction in the quasi-Bragg regime is illustrated in fig.~\ref{figRabi}. We have performed measurements of the populations in the different momentum states after a pulse of order $n=3$ with a peak two-photon Rabi frequency $\gamma_{max}=3.3$. The pulse duration $\sigma$ is scanned between $0.05\omega_r^{-1}$ and $1.1\omega_r^{-1}$. The distribution of the population over the different momentum states is well reproduced by the numerical simulation including the velocity selection effect for long pulses. The very good agreement between the data and the simulation does not rely on any data adjustment. The simulation only needs two parameters which have been independently measured: the atomic velocity dispersion $\sigma_{\tilde{v}}$ has been measured by time-of-flight and the optical power $P$ allowing a calculation of the two-photon Rabi frequency $\gamma_{\mathrm{max}}$ (see appendix~\ref{ap1}). In addition the calculated $\gamma_{\mathrm{max}}$ is compared to first order Rabi oscillation with rectangular pulses.

The SP and LP regimes are still distinguishable with a finite velocity dispersion ensemble. Landau-Zener losses are significant for short pulses ($\sigma < 0.4\omega_r^{-1}$) and completely vanish for longer pulses. The velocity dispersion has a major impact in the LP regime reducing the oscillation amplitude of the populations of the Bragg states as the spectral width of the pulse becomes narrower. In order to minimize the velocity selection and to address all the atomic velocity distribution, one would like to use a pulse as short as possible. However a very short pulse induces Landau-Zener losses and associated diffraction phases which degrade the interferometer performances. Therefore, a trade-off between velocity selection and Landau-Zener losses is needed. It is highly dependent on the shape of the lattice pulse and on the velocity dispersion of the atomic source. The Gaussian shape pulses studied here are commonly used to reduce the spectral width of the pulse and so the Landau-Zener losses, but it does not take the full advantage of the available laser power to drive Rabi oscillations. One could think to use smooth edge-functions with a larger pulse area \cite{Gochnauer2019} for the same $\gamma_{\mathrm{max}}$ value to manage a better trade-off: an example of such a pulse is the "tanh-pulse". A lower velocity dispersion, achieved by velocity selection or delta-kick collimation techniques \cite{Muntinga2013}, gives more room between short pulses Landau-Zener losses and long pulses spectral narrowing. 

The limited laser power and the finite velocity dispersion allow us to achieve a mirror pulse efficiency above $70\%$ up to the order n = 6 (see fig.~\ref{CvgRabi6}). Such an efficiency ensures a good fringe visibility for three-pulse interferometers. However, it corresponds to pulse durations in the SP regime where losses are not negligible and diffraction phases are not trivial and hard to control to a metrological level. 

\begin{figure}
\centerline{\includegraphics[width=0.5\textwidth]{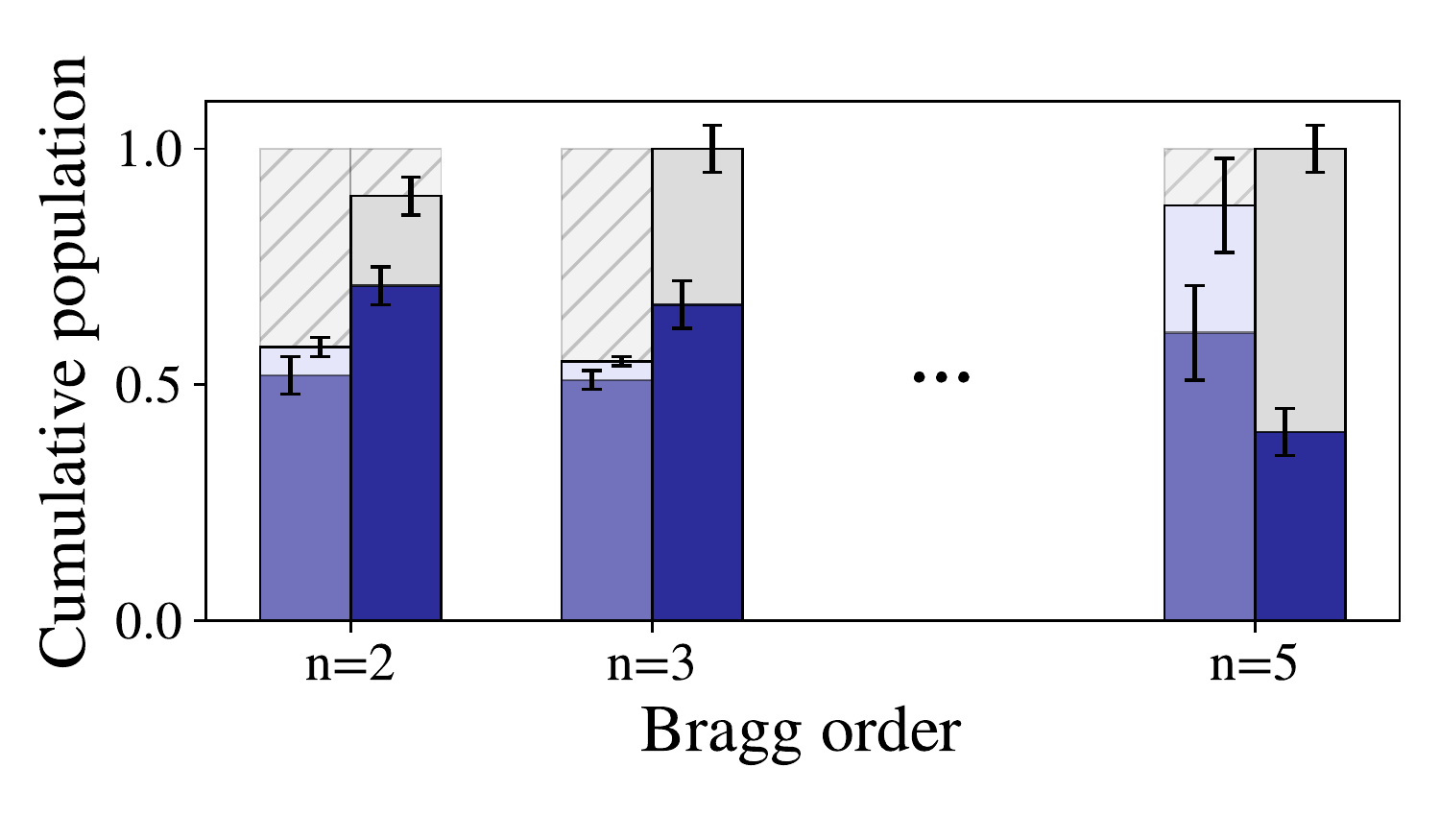}}
\caption{(colors online) Population distribution after a mirror pulse in the SP regime (left columns) and in the LP regime (right columns). The dark (resp. light) color bar area stands for the transferred fraction $P_n$ in $\ket{2n \hbar k}$ (resp. non-transferred fraction $P_0$ in $\ket{0 \hbar k}$). The hatched areas quantify the fraction of atoms lost in the unwanted states. The error bars are standard deviations over 10 realizations. The cumulative height of the two colored areas indicates the fraction of atoms detected and used for the calculation of the relative population at the output of the interferometer. The efficiencies restricted to the two-Bragg states $\frac{P_n}{P_0+P_n}$ for the SP pulses (resp. LP) are $90\%$ ($79\%$) for $n=2$, $93\%$ ($67\%$) for $n=3$ and $70\%$ ($40\%$) for $n=5$.}
\label{figLPvsSP}
\end{figure}

In order to study the impact of this choice in the quasi-Bragg regime on the interferometer signal we first compare the performances of individual mirror pulses in the SP or LP regime for Bragg order up to $n=5$. Figure \ref{figLPvsSP} shows how the population distributes over the different momentum states, including unwanted states, after the mirror pulses. Here, the LP pulses correspond to $\theta_\mathrm{R}=3\pi$ that match the compromises illustrated in fig.~\ref{sigop}, while the SP pulses correspond to $\theta_\mathrm{R}=\pi$ for the same $\gamma_\mathrm{max}$. For example, in the $n=3$ case illustrated in fig.~\ref{figRabi}, the LP mirror pulse duration is $\sigma\simeq 0.42\omega_r^{-1}$ and the SP mirror pulse duration is $\sigma\simeq 0.10\omega_r^{-1}$. 

The LP pulses conserve the major part of the population in the two Bragg states but the transfer efficiency is limited and decreases as the velocity selectivity increases with the Bragg order. On the contrary, the SP pulses perform high efficiencies restricted to the two-Bragg states $\frac{P_n}{P_0+P_n}$, despite a significant loss of atoms in the unwanted states. These two different kinds of losses have different impacts on the interferometer signal that will be discussed in the next section. The drop in transfer efficiency for $n=5$ is due to the limited laser power that prevents us to find the optimal pulse parameters for such a high-order. 

%%%%%%%%%%%%%%%%%%%%%%%%%%%%%%%
\subsection{Quasi-Bragg Atom interferometer}

We perform Mach-Zehnder type three-pulse interferometers. In the usual two-level picture, relevant for Raman pulses or in the deep Bragg regime, the beam-splitter pulses correspond to $\theta_{\mathrm{R}}=\pi/2$ and the central mirror pulse corresponds to $\theta_{\mathrm{R}}=\pi$. However, in the quasi-Bragg regime, such low Rabi phases would have been obtained for short pulses leading to large Landau-Zener losses. In the following experiments, the beam-splitters are realized with longer pulses giving $\theta_{\mathrm{R}}=3\pi/2$ (see fig~\ref{rabi}.a). It corresponds to a pulse at the frontier between the SP and the LP regimes and thus represents a compromise between Landau-Zener losses and velocity selection.

\begin{figure}
\centerline{\includegraphics[width=0.45\textwidth]{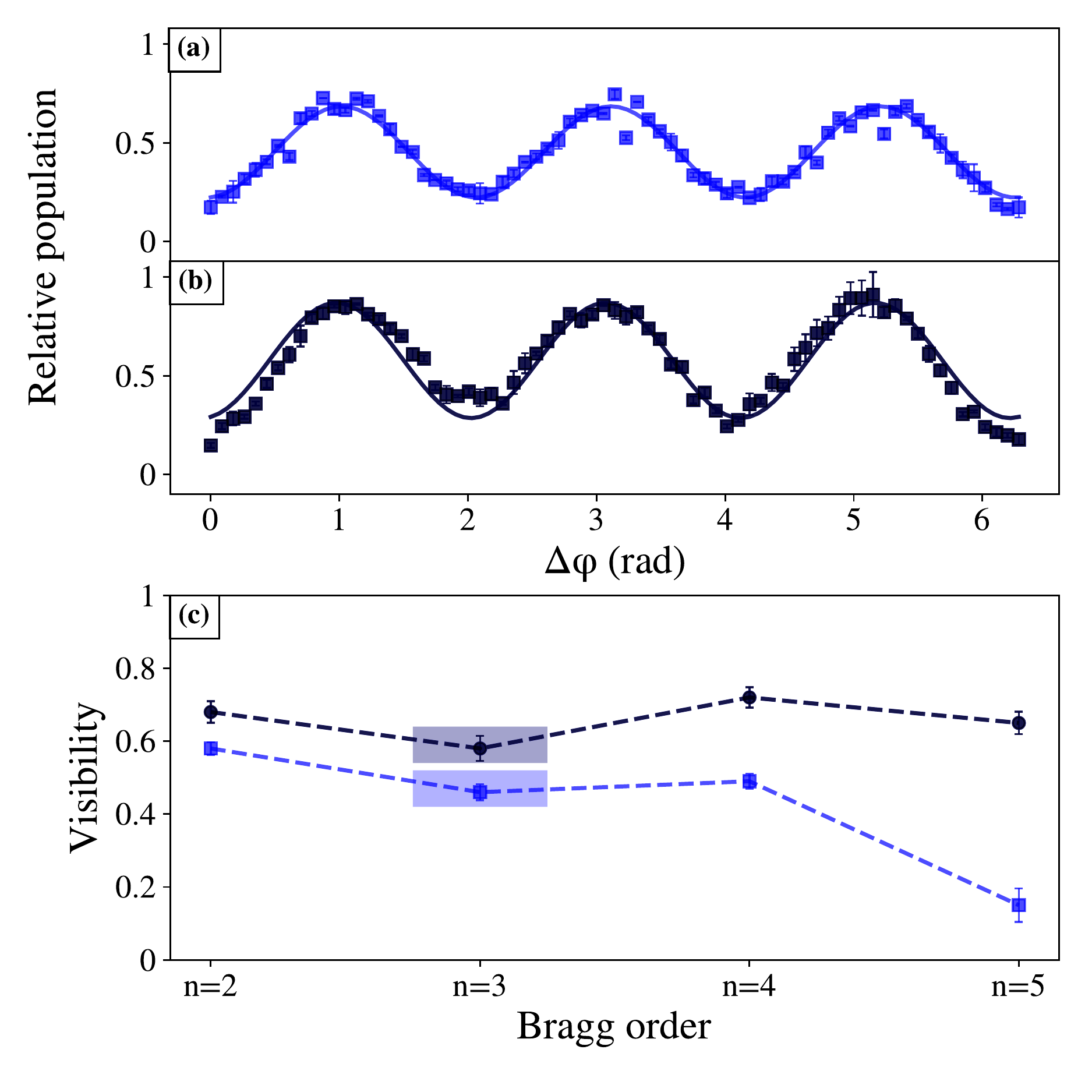}}
\caption{(colors online) Sample interference patterns for mirror pulses in the LP (a) and SP regime (b) for $n=3$. We show the relative atom number in one of the two main output ports normalized by the population in both main ports as a function of the phase shift. Error bars are standard deviations over 3 realizations. The solid blue lines are sinusoidal fits to the data. (c) Fitted visibility in the SP regime (black circles) and in the LP regime (blue squares) as a function of the Bragg order. Error bars for SP interferometers reflect the distortion of the signal and the mismatch with the pure sine model.}
\label{figFringes}
\end{figure}

We build interferometers up to $n=5$ with the different central mirror pulses studied in fig.~\ref{figLPvsSP}. The interferometer phase is scanned by adding a phase shift $\Delta\varphi$ to one of the laser beams forming the lattice prior to the last Bragg-pulse. Figures~\ref{figFringes}.a and b show, as an example, the interference fringes for $n = 3$ for the two configurations of the central pulse. The population in one of the output port is normalized to the population in the two main output ports and is fitted with a sine function as expected for a two paths interferometer. The interrogation time between two pulses is small ($T= 1~\mathrm{ms}$) in order to minimize the effect of environmental perturbations, such as vibrations, and so to focus on the impact of the diffraction losses on the interferometer signal.  In addition, for interferometers with large arms separations, the atoms experience different optical fields on each arm. Therefore, the absolute AC-Stark shift averages unevenly on each arm, which reduces the visibility. However, for the interrogation time $T$ considered in this paper, the separation is small enough to ignore this effect. In addition, for larger arm separation, it is possible \cite{Kovachy15} to compensate the mean AC-Stark shift by using a laser with a supplementary frequency corresponding to a detuning of an opposite sign than the Bragg diffraction beams. The relative intensity of this supplementary frequency beam must be adjusted with respect to the Bragg beams so to cancel the mean AC Stark shift.

In the LP regime (see fig.~\ref{figFringes}.a for $n=3$), the fringe pattern stays very close to a pure sine. It indicates that all atoms are measured in one of the two Bragg states with vanishing population in the unwanted states. The fitted visibilities for LP interferometers are plotted as blue points in fig.~\ref{figFringes}.c. The fringe visibility decreases with the Bragg order $n$ as the velocity bandwidth of the diffraction pulse decreases with $n$, so that a larger fraction of the population remains in the initial state and does not contribute to the fringe signal. The finite available laser power limits the Bragg order to $n=5$ with a fitted visibility around $15\%$.

The SP regime offers the possibility of addressing a broader velocity distribution which contributes to keep an almost constant visibility up to order $n=4$ (see grey points in fig.~\ref{figFringes}.c). The degraded performance for $n=5$ is also due to limited laser power. The improved visibility is explained by the inefficiency of the pulses which populates unwanted momentum states that are filtered out by the normalization and do not enter in the estimation of the plotted relative population. Therefore, the total number of atoms contributing to the interferometer signal is reduced with respect to the LP counterpart (see fig.~\ref{figLPvsSP}). In addition, the population in the unwanted states leads to large diffraction phases and parasitic interferometric paths that distort the fringe patterns. In this regime, the interferometer is no longer a two-path interferometer leading to a degraded phase estimation (see fig.~\ref{figFringes}.b).

\begin{figure}
\centerline{\includegraphics[width=0.5\textwidth]{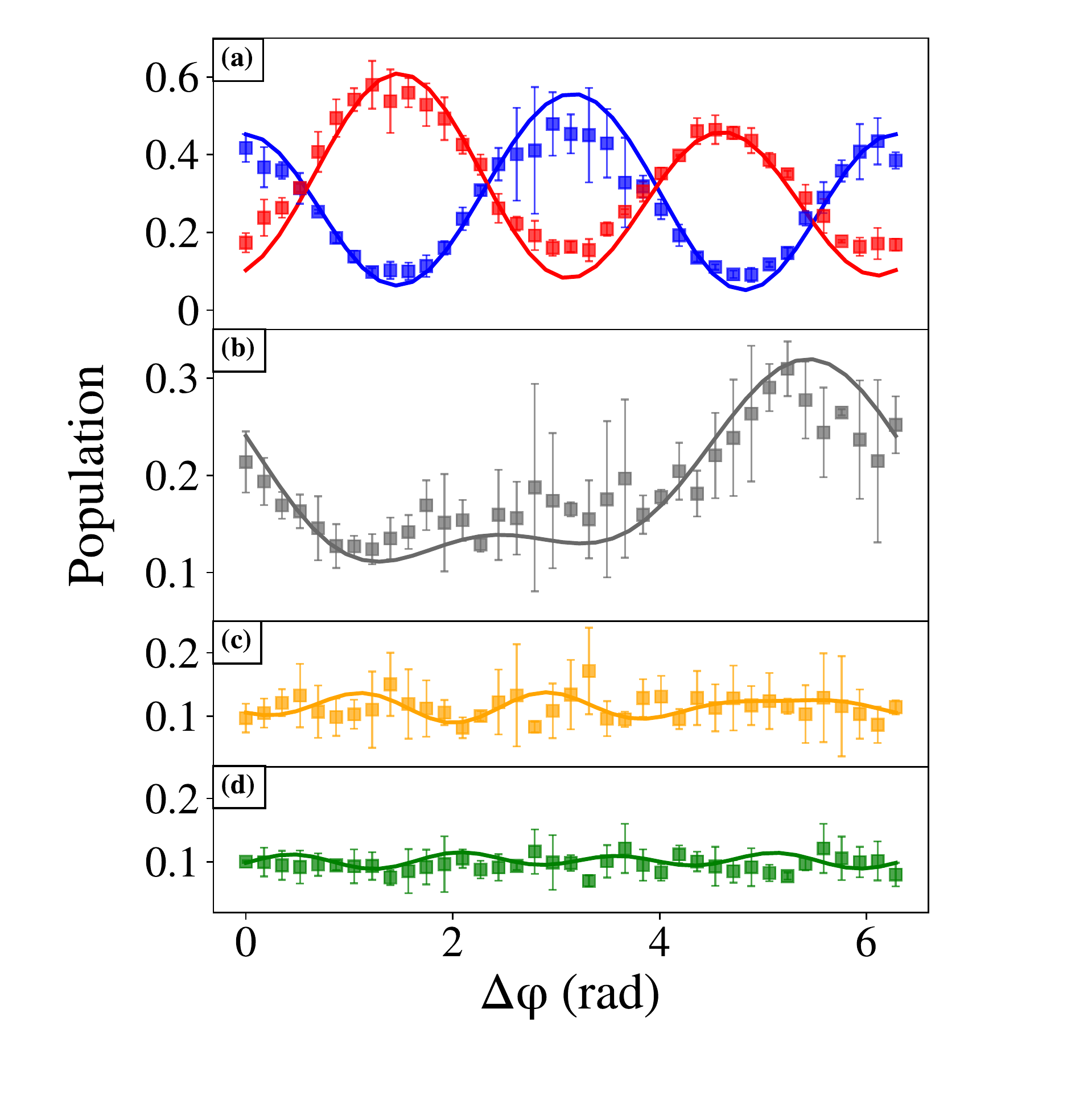}}
\caption{(colors online) Fringe patterns for $n = 2$. The pulses parameters are $\gamma_{\mathrm{max}}=1.75$, $\sigma=0.46\omega_r^{-1}$ for the beam-splitter pulses ($\theta_\mathrm{R}=5\pi/2$) and $\sigma=0.15\omega_r^{-1}$ for the mirror pulse ($\theta_\mathrm{R} \sim\pi$). The multi-port nature of Bragg diffraction leads to distorted fringe patterns. Momentum states depicted are respectively: (a) the two Bragg states, (b) the inner state, (c) and (d) the two outer states. Each experimental data is an average of three points and the error bar is the standard deviation. Solid lines are results of numerical simulation for a velocity dispersion $0.32 v_r$.}
\label{figInterf}
\end{figure}

The multi-port nature of the interferometer in the quasi-Bragg regime plays an already significant role for order $n = 2$ as shown in fig.~\ref{figInterf}. It shows the populations in all the different measurable output ports normalized to the total number of detected atoms. The experimental data are compared with our numerical calculation.

As expected for $n=2$ order pulses, the two main output ports oscillate with a period $\pi$ as the laser phase shift $\Delta\varphi$ is imprinted twice during the diffraction event (see fig.~\ref{figInterf}.a). The fringe pattern is strongly distorted and this distortion is obviously related to the strong modulation of the population of the inner state which is unique in the $n = 2$ case. This behavior is a typical signature of interferometers involving more than two paths and leads to a systematic error of the phase measurement since the response function strongly differs from the expected sine shape. The populations in the outer states are non negligible, but their modulation is much weaker and they do not have a significant impact on the phase measurement for the present experimental parameters.

Spurious interferometers have been observed in dual interferometers from the residual errors of a Lissajou fitting \cite{Parker2016}. In particular, the authors observed a "magic" pulse duration minimizing the errors on the phase extraction. We think the effect is related to a minimum of unwanted momentum states populated during the $\pi/2$-pulses in the SP regime. We believe that our direct observations of the fringes on each interferometers outputs will guide the theories needed for reading the phase taking into account multi-port features \cite{Gaaloul}.

%%%%%%%%%%%%%%%%%%%%%%%%%%%%%%%
%%%%%%%%%%%%%%%%%%%%%%%%%%%%%%%
\section{Conclusions}
We have investigated atom diffraction and interferometry of a BEC by a standing light wave in the quasi-Bragg regime up to the sixth Bragg order. We have shown simulations, with no adjusted parameters, in good agreement with experimental results for arbitrary pulse length regimes (Short and Long Pulses) and initial velocity dispersion. In particular, we have modeled with very good agreement with the experiment all the momentum states involved in the diffraction process: Bragg states and non adiabatic losses. Our work points out the required size of the momentum basis for accurate calculations of diffraction process. By matching simulations and experiments we provide a quantitative understanding of the inherent diffraction phase-shifts and multi-port features of LMT interferometers based on quasi-Bragg diffraction. We confirmed the link between diffraction phases and the non adiabatic losses and our simulations anticipate diffraction phase shifts of up to several tens of mrad. We demonstrated that an optimal diffraction efficiency at the border between the SP and the LP regime for a Rabi phase $\theta_{\mathrm{R}} \ge \pi$.  Higher Rabi phase, i.e. higher $(\sigma,\gamma_{\mathrm{max}})_{\rm{opt}}$, would lead to a similar diffraction efficiency limited by velocity selection and a reduced diffraction phase, but might impact effects scaling in $\gamma_{\mathrm{max}} \sigma$ such as the AC-Stark shift or spontaneous emissions. We have also demonstrated interferometer fringes with phase-controlled interferometers up to n=5. Therefore, we were able to observe the impact of the multi-path interferences directly on the fringe's signal for each interferometer outputs. The insight gained from these investigations should guide the development of new methods for better optimizations and estimations of the multi-port interferometers.

%%%%%%%%%%%%%%%%%%%%%%%%%%%%%%%
%%%%%%%%%%%%%%%%%%%%%%%%%%%%%%%
\section{Acknowledgments}
We acknowledge the contributions made by the past members (B. Decamps, J. Alibert, M. Bordoux). We thank  J.P. Gauyacq, N. Gaaloul, K. Hammerer and J.-N. Siem{\ss} for fruitful discussions. This research was supported by the research funding grant No. ANR-19-CE47-0002.

%%%%%%%%%%%%%%%%%%%%%%%%%%%%%%%
%%%%%%%%%%%%%%%%%%%%%%%%%%%%%%%
\appendix
%%%%%%%%%%%%%%%%%%%%%%%%%%%%%%%
%%%%%%%%%%%%%%%%%%%%%%%%%%%%%%%

\section{Lattice depth measurement $\Omega$}
\label{ap1}
Bragg scattering can be described as a multiphoton Raman process between momentum states. Each elementary 2-photon Raman process couples momentum states in the same internal ground states $\ket{F_g,m_{F_g}}$ through an intermediate excited state in a different internal state $\ket{F_e,m_{F_e}}$. The single-photon Rabi frequencies between a ground state and an intermediate state is denoted $\Omega_{\ket{e}}$. The excited states have a lifetime $1/\Gamma$. For a large detuning ($\Delta \gg \Omega_{\ket{e}}, \Gamma$) from intermediate level $\ket{e}$ (see figure \ref{parabol}), the system can be described as an effective two-level system by adiabatically eliminating the intermediate level $\ket{e}$. The effective Rabi frequency between two momentum states is given by summing over all possible intermediate states:
\begin{equation}
\Omega = \sum_{\ket{e}} \frac{\Omega_{\ket{e}}^2}{2 \Delta}
\label{Rabi2photon}
\end{equation}
In the electric-dipole approximation, $\Omega_{\ket{e}}$ is proportional to the product of the electric field amplitude by the matrix element of the dipole operator. Using the Wigner-Eckart theorem we factor out the reduced matrix element and Clebsch-Gordan coefficient. 
\begin{eqnarray}
\Omega_{\ket{e}} = E_0 \sqrt{\frac{3\epsilon_0 \lambda^3 \Gamma}{2 \pi^2 \hbar}} \sqrt{(2 F_e+1) (2 F_g+1)} \nonumber \\[0.5 cm]
\times \begin{Bmatrix}
J_e & 1 & J_g\\
F_g & I & F_e
\end{Bmatrix}
\begin{pmatrix}
F_e & 1 & F_g\\
-m_{F_e} & q & m_{F_g}
\end{pmatrix}
\end{eqnarray}
In practice, we use the $D_2$ lines of the $^{87} Rb$, corresponding to a wavelength $\lambda = 780.1$ nm, and a natural line-width $\Gamma = 2\pi \times 6.1$ MHz. $\epsilon_0$ is the vacuum permittivity. The laser field polarization is given in standard components $q = \pm 1, 0$. The amplitude of the electric field corresponding to a single photon transition is evaluated from the power $P$ and the Gaussian beam waist $w_0$:
\begin{equation}
E_0 = \frac{2}{w_0}\sqrt{\frac{P}{\pi c \epsilon_0}}
\end{equation}
In our experimental setup, the retro-reflected configuration leads to two optical lattices with orthogonal polarization. The optical power for each lattice's arm is recycled by the retro-reflection. Therefore, $P$ is the optical power measured in the incident laser beam for each frequency $\omega_1$ or $\omega_2$ (\textit{i.e.} the total incident optical power is $2P$). 

All the data reported in this paper are performed with atoms prepared in the ground state $(J_g =1/2$, $F_g =1$, $m_{F_g} = 0)$ and the optical lattice with the circular polarization $\sigma_-$ ($q=-1$ in standard components). Therefore, only two intermediate states $\ket e$ are coupled: $J_e =1/2$, $F_e =1,2$, $m_{F_e} =-1$.

\begin{figure}
\centerline{\includegraphics[width=0.5\textwidth]{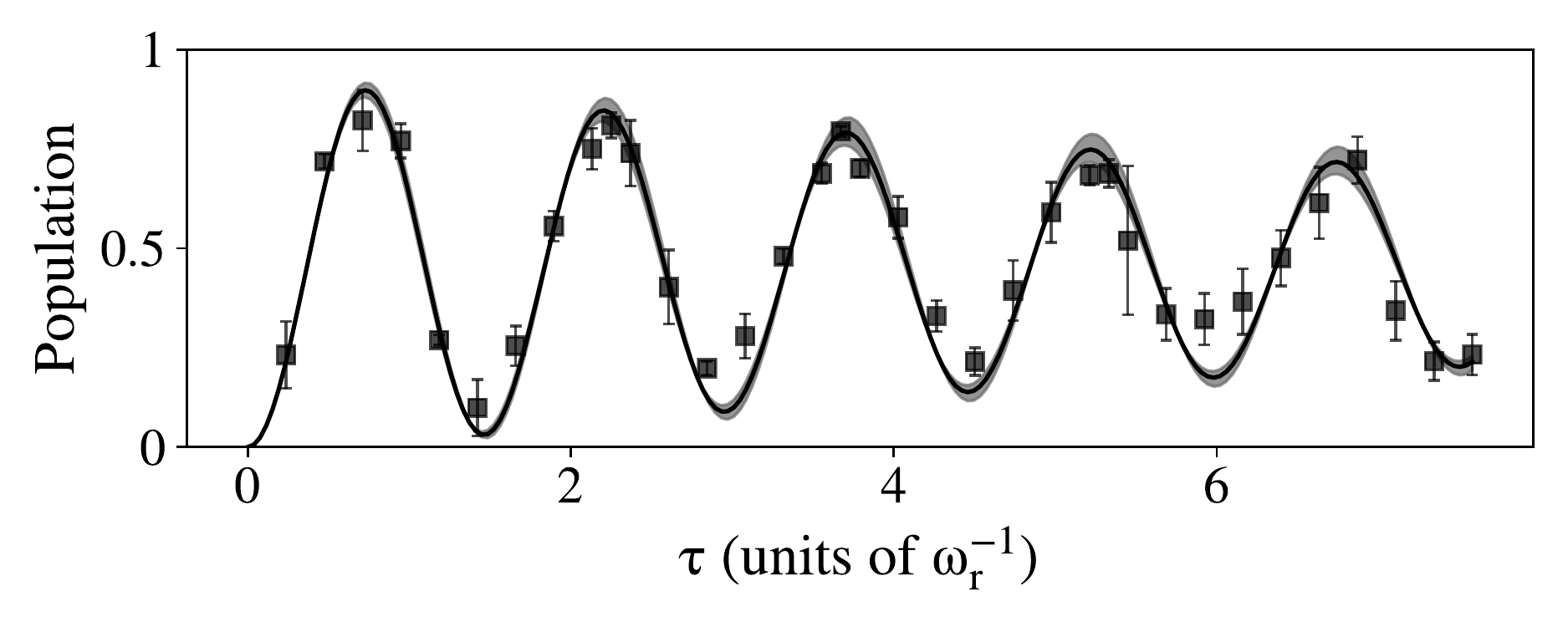}}
\caption{Evolution of the diffracted Bragg state for a resonance condition for Bragg order $n = 1$. Temporal of the envelope of the lattice is rectangular.}
\label{figRabiN1}
\end{figure}

The effective two photon Rabi frequency (eq.~\ref{Rabi2photon}) is a central parameter in the atom diffraction with an optical lattice. In order to determine $\Omega$, we measured the Rabi oscillations between the two diffracted states with a rectangular pulse 1st order Bragg diffraction in the Bragg regime Figure~\ref{figRabiN1}. We obtain a good agreement between the measured and calculated $\Omega$.

\section{Truncation of the momentum basis used for the numerical model}
\label{ap3}

\begin{figure}
\centerline{\includegraphics[width=0.5\textwidth]{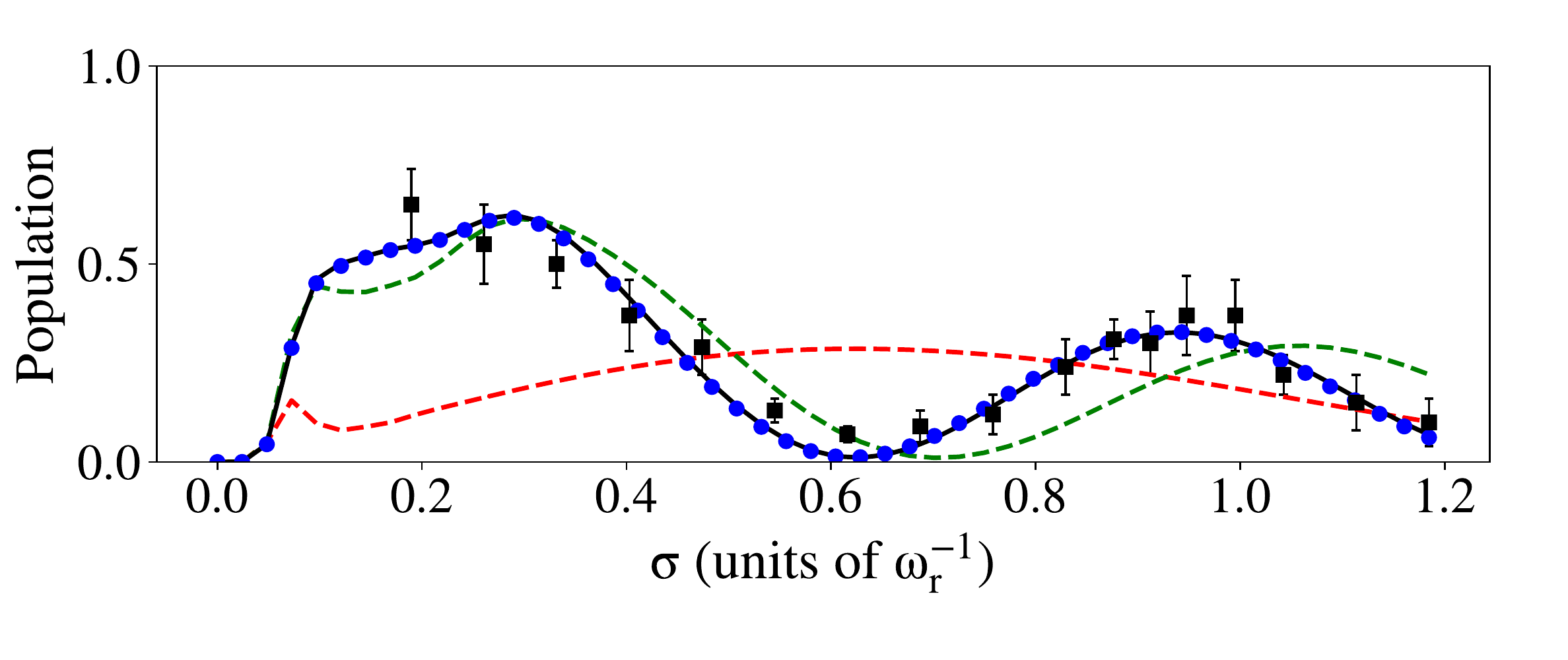}}
\caption{(colors online) Population in $\ket{12\hbar k}$ state for $n=6$ and $\gamma_{\mathrm{max}}=6.25$ as a function of the pulse duration. Black squares are the experimental measurements. Error bars are standard deviations on 10 measurements. The experimental data are compared with numerical simulation results for the corresponding pulse parameters and atomic velocity dispersion for various $m$ values. The dashed lines are simulation results for $m=0$ (red) and $m=1$ (green). The black solid line is the result for $m=2$ and overlaps with blue dotted line corresponding to $m=3$.} 
\label{CvgRabi6}
\end{figure}

We have shown in fig.~\ref{gauss} that the dynamics of high-order Bragg diffraction requires to consider a sufficiently large momentum basis.  For example, we compare the measured diffraction probability for a quasi-Bragg order $n = 6$ to its numerically predicted value for different basis truncations.  Figure~\ref{CvgRabi6} presents this comparison for a large range of the pulse duration $\sigma$. For $m=0$, meaning that the basis only contains the two Bragg states and the inner states in between, the numerical result fails estimating both the amplitude and the frequency of the population oscillation. Considering two outer states ($m=1$) increases the agreement with the experimental results. The complex dynamics is properly captured for $m\geq2$. The overlap of the two curves for $m=2$ and $m=3$ confirms the convergence of the simulation. 

We numerically solve the system of differential equations using the Scipy.integrate package of Python\texttrademark. The computational cost rapidly increases  with the basis size, especially when considering a non-zero velocity dispersion. It is thus interesting to find a criterion giving the minimal value $m_0$ of outer states that has to be considered for a given quasi-Bragg order $n$.

A first naive energetic criterion is related to the fact that every states $\ket{2l\hbar k}$ in between the two Bragg states ($l\in [0,n]$) are naturally included in the calculation. The model should then at least consider outer states down to $\ket{-2(m_0) \hbar k}$ so that $\delta_{-m_0} \geq \max_{l\in [0,n]} |\delta_l|$. Note that the exact same criterion is obtained considering outer states up to $\ket{2(n+m_0) \hbar k}$ since the diagonal term $\delta_l$ is symmetric. When the Bragg condition is realised for the quasi-Bragg order $n$, the maximum detuning for inner states reads:
\begin{eqnarray}
\max_{l\in [0,n]} |\delta_l| = 
 \left\{
\begin{array}{l}
|\delta_{n/2}|=n^2/4\mathrm{\quad if\ \textit{n}\ is\ even,} \\
|\delta_{(n-1)/2}|=(n^2-1)/4\mathrm{\quad if\ \textit{n}\ is\ odd.} \end{array}
\right. 
\end{eqnarray}

Solving the equality and rounding to the next integer gives the criterion on $m_0$:
\begin{eqnarray}
m_0 = 
\left\{
\begin{array}{l}
\left\lceil\frac{1}{2}n\left(\sqrt{2} -1\right)\right\rceil\mathrm{\quad if\ \textit{n}\ is\ even,} \\
\\
\left\lceil\frac{1}{2}n\left(\sqrt{2-\frac{1}{n^2}} -1\right)\right\rceil\mathrm{\quad if\ \textit{n}\ is\ odd,} \end{array}
\right. 
\end{eqnarray}
where $\lceil x \rceil$ is the ceiling function that rounds $x$ to the least integer greater than or equal to $x$. 

In practice, for $n>3$, the $1/n^2$ term for odd quasi-Bragg order does not affect the rounded value for $m_0$. We thus take $m_0=\left\lceil\frac{1}{2}n\left(\sqrt{2} -1\right)\right\rceil$ as a lower bound for $m$.

\begin{figure}
\centerline{\includegraphics[width=0.5\textwidth]{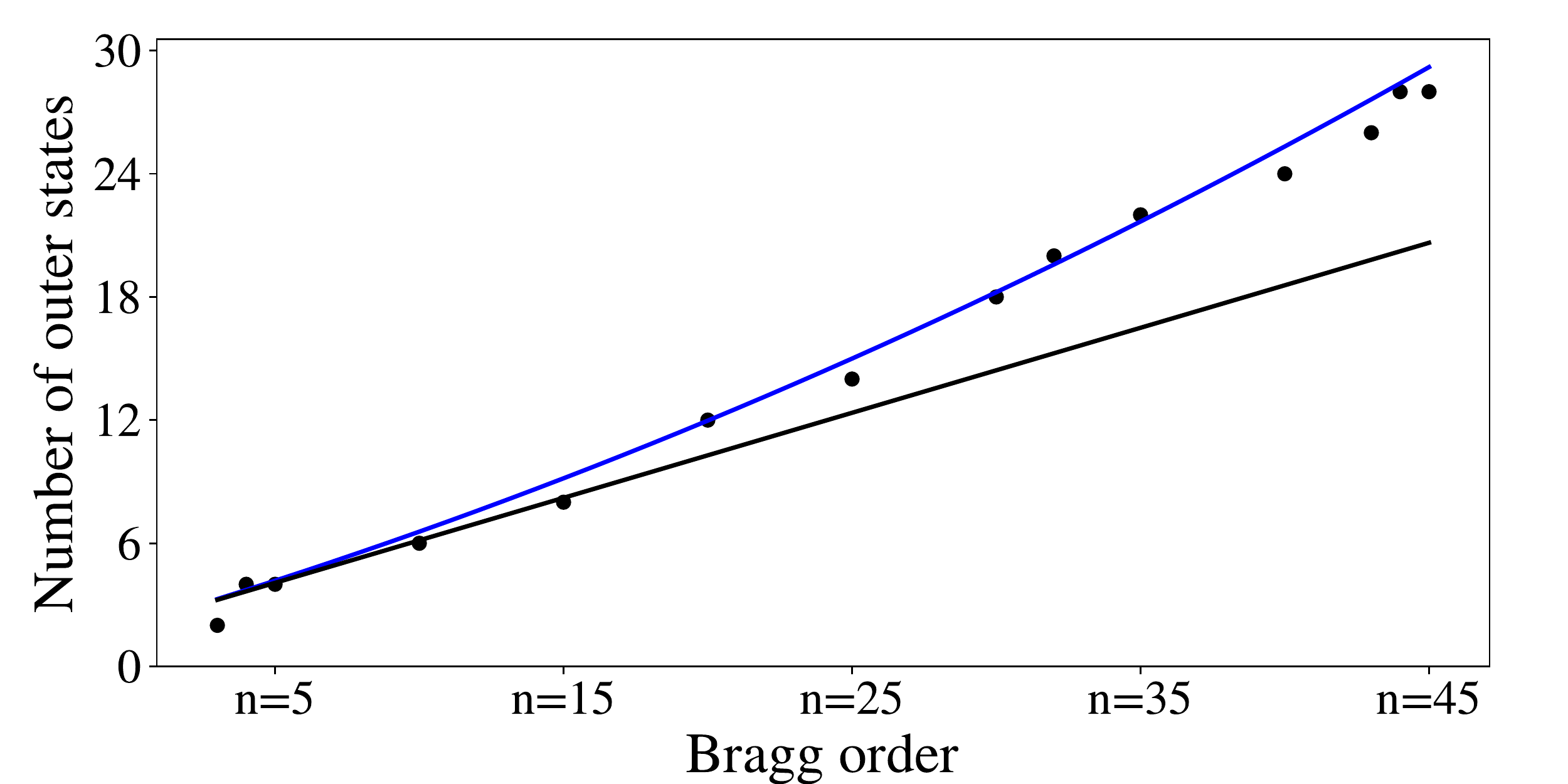}}
\caption{(colors online) Comparison of the minimum required number of outer states as a function of the quasi-Bragg order $n$. Black dots are results of a convergence test $2m_c$. The solid black line is the lower bound criterion $2m_0$ derived from the energetic argument. The blue solid line is an empirical correction to the criterion adding a $n^2$ contribution (eq.~\ref{m0}). The lines correspond to the non-rounded analytical formulas for $2m_0$ and $2m_0^{emp}$.}
\label{stairs}
\end{figure}

This criterion is compared to a convergence test of the truncation of the calculation basis performed in the case of zero velocity dispersion. The quantity of interest is the peak Rabi frequency $\gamma_{\mathrm{max}}$ that gives a perfect mirror pulse (diffraction probability $>99\%$) with the shortest pulse length. This quantity varies with $m$ and converges to a constant value for $m>m_c$. Figure~\ref{stairs} shows the comparison between the values of $m_c$ (black dots) and the previously derived criterion $m_0$ (black line) for different quasi-Bragg order $n$.

The minimum required numbers of outer states is well predicted by the criterion $m_0$ up to a quasi-Bragg order $n\simeq10$. It appears that it does not capture the $n^2$ term relevant for high $n$ \cite{SzigetiNJP12}. We propose a modified criterion $m_0^{emp}$ including a minimal fitted $n^2$ term that reproduces the convergence test results:
\begin{equation}
  m_0^{emp}=\left\lceil\frac{1}{2}n\left(\sqrt{2}-1\right)+1+An^2\right\rceil, 
\label{m0}
\end{equation}
where $A \approx 2 \times 10^{-3}$.

This analytical criterion is based on a simple energetic argument and compared to zero-velocity calculations. It gives a lower bound of the number of outer states that has to be included to accurately reproduce the dynamics. For relevant experimental parameters, with large velocity dispersion and/or high peak two-photon Rabi frequency, one might have to run a dedicated convergence test to ensure that all the relevant off-resonant couplings are taken into account. 

%------------------------------------
%------------------------------------
%\bibliographystyle{apsrev4-1} 

%\bibliography{reference} 


\end{document}